\documentclass[trackchanges,twocolumn]{aastex701}

\usepackage{graphicx}
\usepackage{natbib}
\usepackage{multirow}
\usepackage{amsmath}
\usepackage{booktabs}
\usepackage{tablefootnote}
\usepackage{threeparttable}
\usepackage{appendix}
\begin{document}

\title{Multi-thermal Structure of Solar Jets based on Combined EUV and H$\mathrm{\alpha}$ Observations}

\author[orcid=0009-0003-4609-3177,sname='Lei']{Lei Huang}
\affiliation{School of Astronomy and Space Science, Nanjing University, Nanjing 210023, People’s Republic of China}
\affiliation{Key Laboratory of Modern Astronomy and Astrophysics (Nanjing University), Ministry of Education, Nanjing 210023, People’s Republic of China}
\email{huanglei@smail.nju.edu.cn}  

\author[orcid=0000-0002-9293-8439,sname='Yang']{Yang Guo} 
\affiliation{School of Astronomy and Space Science, Nanjing University, Nanjing 210023, People’s Republic of China}
\affiliation{Key Laboratory of Modern Astronomy and Astrophysics (Nanjing University), Ministry of Education, Nanjing 210023, People’s Republic of China}
\email[show]{guoyang@nju.edu.cn}

\author[orcid=0009-0007-8418-1986,sname='Zining']{Zining Ren} 
\affiliation{School of Astronomy and Space Science, Nanjing University, Nanjing 210023, People’s Republic of China}
\affiliation{Key Laboratory of Modern Astronomy and Astrophysics (Nanjing University), Ministry of Education, Nanjing 210023, People’s Republic of China}
\email{rzn@smail.nju.edu.cn}

\author[orcid=0000-0002-4978-4972,sname='Mingde']{Mingde Ding} 
\affiliation{School of Astronomy and Space Science, Nanjing University, Nanjing 210023, People’s Republic of China}
\affiliation{Key Laboratory of Modern Astronomy and Astrophysics (Nanjing University), Ministry of Education, Nanjing 210023, People’s Republic of China}
\email{dmd@nju.edu.cn}

\begin{abstract}
This study investigates the multi-thermal structure of solar jets by combining extreme ultra-violet (EUV) and H$\alpha$ observations, addressing whether cool and hot components originate from the same jet, their morphological similarities and differences, and the role of mini‑filaments. We analyze 70 jet events simultaneously observed by Atmospheric Imaging Assembly (AIA) and Chinese H$\alpha$ Solar Explorer (CHASE) from 2023 May 25 to 2026 January 15. The events are classified into four types based on whether or not there are significant differences between the two wavebands in terms of length and width; nonlinear force‑free field extrapolations are applied to examine magnetic fields. We find that the cool and hot components correspond to different temperature parts of a unified jet, showing spatiotemporal coincidence and kinematic synchronization. Morphologically, 30\% show no significant difference in length/width, 37\% differ in one dimension, and 33\% differ in both. About 80\% involve mini‑filament eruptions; jets with filaments more often exhibit morphological differences (filament non-expansion, hot shell thickness, insufficient kinetic energy, or material rarefaction). In contrast, jets without filaments tend to show consistent morphologies; when discrepancies occur, they are mainly linked to small magnetic field curvature, which reduces magnetic tension and leads to less acceleration of cool material.
These results highlight the complex interaction between thermal, dynamic, and magnetic factors in shaping jet morphology.

\end{abstract}

\keywords{\uat{Solar activity}{1475} -- \uat{Solar filament eruptions}{1981} -- \uat{Solar magnetic reconnection}{1504}}

\section{Introduction}
Jets are prevalent eruptive phenomena on the Sun, typically manifesting as collimated plasma beams (\citealt{Raouafi2016}; \citealt{Shen2021}). They can be categorized by their observational wavebands into H$\alpha$ jets, extreme ultraviolet (EUV) jets, X-ray jets, and white-light jets. H$\alpha$ jets, also known as H$\alpha$ surges, refer to the low-temperature jet component observed in the H$\alpha$ waveband (\citealt{Roy1973}; \citealt{Xu1984}). EUV jets are observed in the extreme ultraviolet wavebands, with temperatures of approximately $10^5$ K to $10^7$ K, representing one of the primary observational forms of coronal jets (\citealt{Schmahl1981}). X-ray jets are high-temperature jets observed in the soft or hard X-ray wavebands, with temperatures reaching 10$^6$ to 10$^7$ K (\citealt{Shibata1992}).

According to existing observational studies, jets exhibit both connections and distinct characteristics across these three temperature ranges. In the higher temperature range, X-ray and EUV jets share similarities in morphology and speed, although EUV jets are generally smaller in scale and shorter-lived (\citealt{Chae1999}). Observations indicate that some X-ray jets are accompanied by coincident H$\alpha$ surges (\citealt{Shibata1992}; \citealt{Shimojo1996}; \citealt{Canfield1996}), while a considerable number of H$\alpha$ surges are not detected with corresponding X-ray emission (\citealt{Schmieder1995}). Regarding the relationship between EUV and H$\alpha$ jets: some studies suggest that they correspond to hot and cool plasma moving along different magnetic field lines, dynamically coupled, but with the cool component often delayed for several minutes (\citealt{Schmieder1994}; \citealt{Alexander1999}). This delay is interpreted as due to the cooling of the hot jet. Structurally, EUV jets exhibit a slightly converging shape with open structures, whereas H$\alpha$ jets are typically smaller in scale, distributed along the edges of the hot jet, and always appear later than their EUV counterparts (\citealt{Jiang2007}). These observations indicate that hot and cool jets may share a common acceleration mechanism, but their formation and evolution involve complex multi-thermal plasma interactions. 

To deeply explore the origins and connections between the cool and hot components in jets, magnetohydrodynamic (MHD) simulations provide critical theoretical insights. Some simulations indicate that both hot and cool jets can be jointly generated by magnetic reconnection driven by emerging flux (\citealt{Nishizuka2008}). Due to the pronounced asymmetry of the reconnection process, low-density regions give rise to high-speed hot jets, while high-density regions produce low-speed cool jets. This naturally explains the observational phenomenon that hot jets often reach greater heights earlier (\citealt{Nishizuka2008}). Specifically, simulations show that when a rising magnetic loop reconnects with the background magnetic field, plasma at the current sheet is heated to the temperature corresponding to X-ray emission and accelerated along field lines, forming hot jets. Meanwhile, cool plasma from the chromosphere is carried upward with the rising loop and ejects at speeds of several tens of kilometers per second under the sling-shot effect, forming H$\alpha$ jets. Magnetic islands containing cool, dense plasma can also form within the current sheet; these islands are ejected horizontally, further contributing to the cool jet component. This physical picture suggests that magnetic reconnection can accelerate cool plasma without significantly heating it, providing a plausible explanation for the coexistence of X-ray and H$\alpha$ jets. Furthermore, simulations show that cool and hot plasmas mix, with an intermediate-temperature plasma around 10$^{5}$ K consistently present between them, which likely corresponds to the intermediate-temperature structures observed in the EUV band (\citealt{Yokoyama1996}). MHD simulations offer plausible physical mechanisms for a unified understanding of the origins, timing, and spatial relationships of multi-waveband jets.

In the exploration of the physical mechanisms of jets, substantial observational evidence suggests that their eruptions are often linked to the activity of mini-filaments. For instance, based on statistics of a series of polar jets, \cite{Sterling2015} note that the vast majority of jets are triggered by accompanying mini-filament eruptions; another study of 23 jets finds that as many as 87\% of the cases show clearly identifiable erupting filaments (\citealt{Baikie2022}). This observational phenomenon is reproduced and validated in numerical simulations. For example, through three-dimensional MHD modeling, \cite{Wyper2017} successfully simulate the physical process wherein a filamentary structure erupts and forms a jet along open magnetic field lines. The presence of filaments provides a crucial clue for explaining the origin of cool plasma in jets. Multiple studies indicate that the cool component of jets likely originates from the mini-filament material erupting at the jet base (\citealt{Moore2010}; \citealt{Shen2012}; \citealt{Wang2018}; \citealt{Chen2020}; \citealt{Sun2023}). Observationally, the visible arch-like filament system can carry cool, dense photospheric or chromospheric plasma upward to coronal heights (\citealt{Bruzek1967}; \citealt{Huang2018}; \citealt{Su2018}). Within the current sheet of magnetic reconnection, this cool plasma can be trapped in the formed magnetic islands and become part of the cool jet as the islands merge and are ejected (\citealt{Yokoyama1995}). Therefore, mini-filament eruptions are not only important dynamic structures triggering jets but also likely the direct carriers of the cool material observed within the jets.

To investigate the multi-thermal configuration of solar jets, we focus on EUV and H$\alpha$ wavebands and systematically compare the hot and cool components by statistically analyzing the morphology of the same jet in both wavelengths. Section \ref{section2} describes the selected events, instruments, and methodology. Section \ref{section3} presents the statistical results and detailed observational analysis of representative cases. Section \ref{section4} provides a summary and discussion of the findings.

\begin{figure*}
   \centering
   \includegraphics[width=0.95\textwidth]{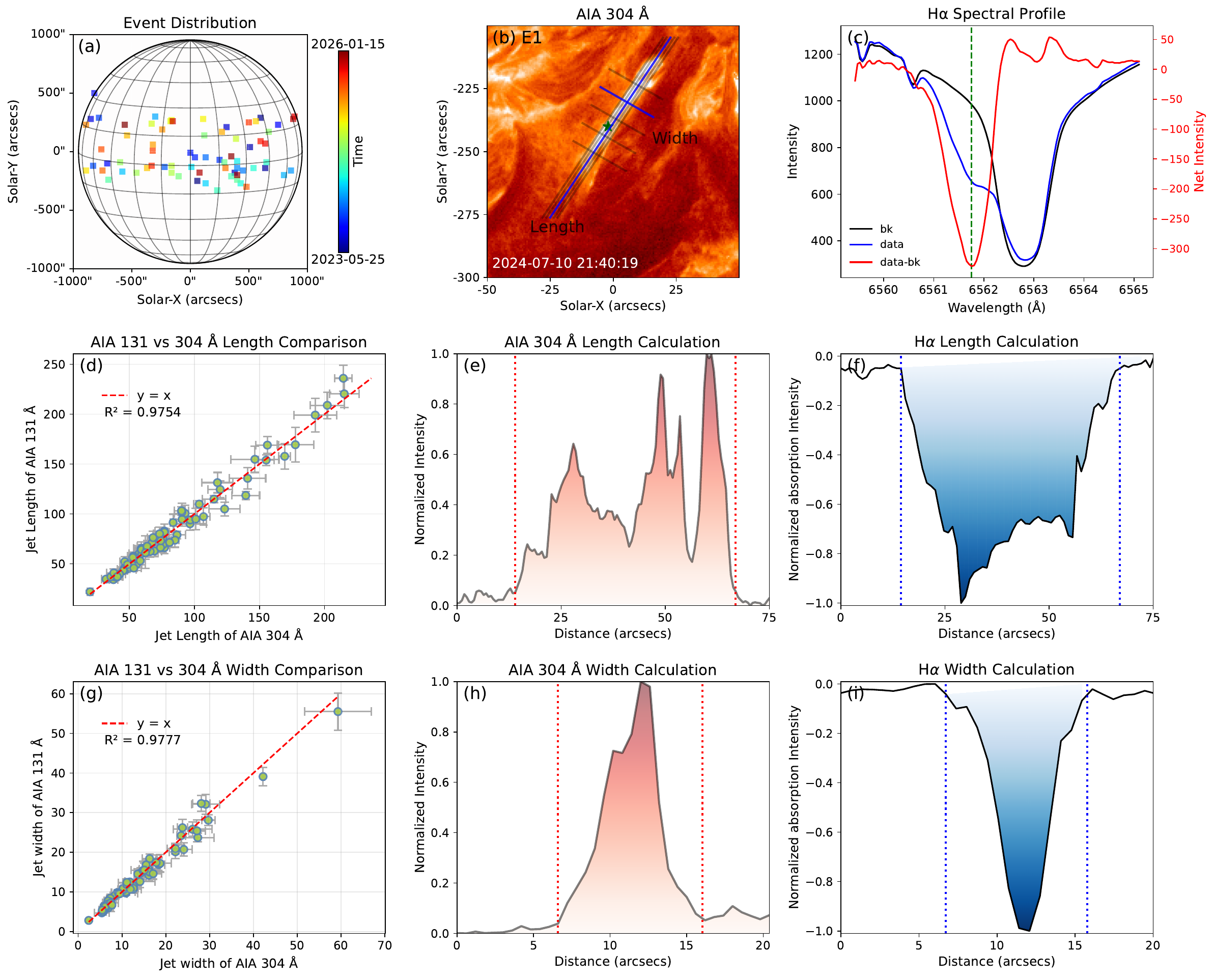}
   \caption{Overview of the statistical method. (a) Time and location of the 70 analyzed jet events. (b) AIA 304 $\mathrm{\AA}$ observation of the jet occurred in 2024 July 10 (hereafter E1), 5 axial and transverse slices are used to calculate its length and width, respectively. (c) CHASE H$\alpha$ spectral line profile at the green point location marked in panel (b). The black line represents the background profile, the blue line shows the original spectral line profile at that point, and the red line is the net profile after background subtraction. The green dashed line marks the wavelength corresponding to the minimum of the net profile. (d) Comparison of the lengths of the 70 jets as measured in the AIA 304 $\mathrm{\AA}$ and 131 $\mathrm{\AA}$ wavebands. (e) Normalized intensity distribution along the axial slice (blue line) in panel (b) in AIA 304 $\mathrm{\AA}$. The segment between the red dashed lines indicates the derived jet length. (f) Normalized distribution of the maximum absorption intensity along the same axial slice from CHASE H$\alpha$, with the jet length defined by the segment between the blue dashed lines. (g)--(i) Similar to (d)--(f), but for the width instead of length.} 
    \label{fig1}
\end{figure*}

\begin{figure*}
   \centering
   \includegraphics[width=0.95\textwidth]{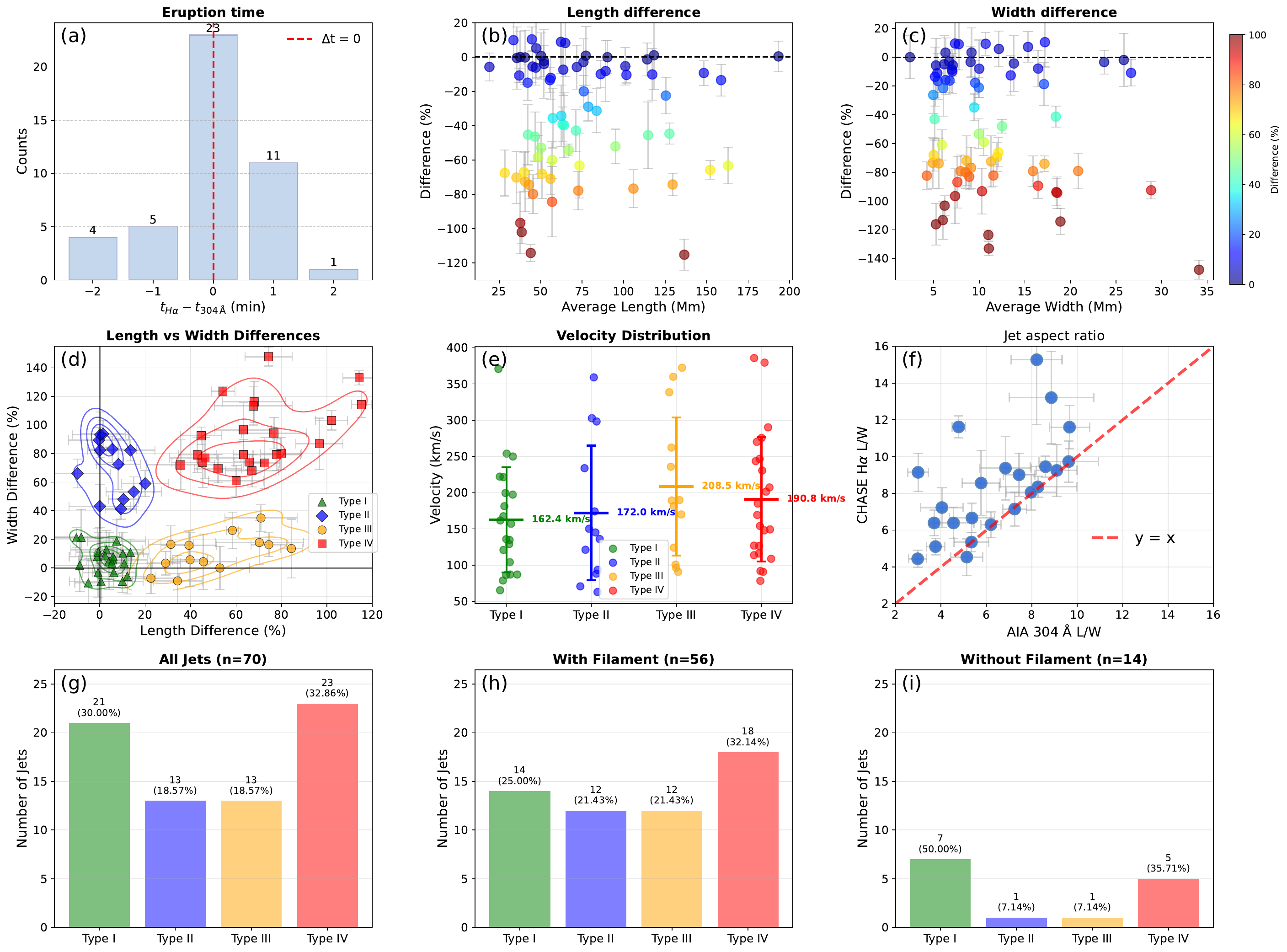}
   \caption{Statistical Results. (a) Distribution of the onset time difference between AIA 304 $\mathrm{\AA}$ and CHASE H$\alpha$ ($\Delta t=t_{\text{H}\alpha}-t_{304\,\text{\AA}}$). (b) Relative length difference between AIA 304 $\mathrm{\AA}$ and CHASE H$\alpha$, defined as $D_L = 2(L_{\text{H}\alpha} - L_{304\,\text{\AA}}) / (L_{\text{H}\alpha} + L_{304\,\text{\AA}})$. (c) Relative width difference between AIA 304 $\mathrm{\AA}$ and H$\alpha$, defined as $D_W = 2(W_{\text{H}\alpha} - W_{304\,\text{\AA}}) / (W_{\text{H}\alpha} + W_{304\,\text{\AA}})$. (d) Length and width differences of the jets, categorized into four types by K-means clustering: Type I - no significant difference in either length or width; Type II - significant difference in width only; Type III - significant difference in length only; Type IV - significant differences in both length and width. The background colored contours represent the results of K-means clustering applied to these jets. (e) Average velocities of the four jet types. (f) Comparison of the length-to-width ratios $L/W$ of Type IV jets in AIA 304 $\mathrm{\AA}$ and in H$\alpha$, the red dashed line indicates $y = x$. (g) Histogram of all the four types of jets (70 events). (h) Histogram of the four types of jets involving mini-filaments (56 events). (i) Histogram of the four types of jets without mini-filament involvement (14 events).} 
    \label{fig2}
\end{figure*}

\section{Observations and Methods}
\label{section2}
We analyze 70 jet events jointly observed by Atmospheric Imaging Assembly (AIA; \citealt{Lemen2012}) aboard Solar Dynamics Observatory (SDO) and Chinese H$\alpha$ Solar Explorer (CHASE; \citealt{Li2022}) from 2023 May 25 to 2026 January 15. Using multi-wavelength observations from AIA, we can obtain images of the jets across various EUV wavebands from cool to hot temperatures. AIA provides images with a pixel size of 0.6$^{\prime\prime}$, a spatial resolution of about 1.5$^{\prime\prime}$, and a temporal cadence of 12 s (\citealt{Lemen2012}). For example, the 131 $\mathrm{\AA}$ channel has contributions from both Fe VIII (characteristic temperature of $10^{5.6}$ K) and Fe XXI ($10^7$ K), so it can simultaneously trace relatively cool and hot plasma. The 171 $\mathrm{\AA}$ channel is dominated by Fe IX, originating from the quiet corona and upper transition region with a characteristic temperature of $10^5$ K. The 304 $\mathrm{\AA}$ channel, primarily emitted by He II, probes the chromosphere and transition region with a characteristic temperature of $10^{4.7}$ K (\citealt{Lemen2012}) and is used to observe low‑temperature plasma structures. These wavebands thus allow us to examine the jet plasma across a broad temperature range. 

CHASE provides full-disk H$\alpha$ spectral observations covering the wavelength range of 6559.7--6565.9 $\mathrm{\AA}$, with a spectral sampling of 0.024 $\mathrm{\AA}$ per pixel and a spatial resolution of about 1.04$^{\prime\prime}$. The H$\alpha$ line is primarily formed in the solar chromosphere and is a key spectral line for diagnosing structures in the lower solar atmosphere, with a characteristic formation temperature of approximately $10^4$ K. The temporal and spatial distributions of the 70 selected jet events are shown in Figure \ref{fig1}(a). 

The selection criteria for the 70 events are as follows. First, we require that each event is clearly visible in both the AIA EUV wavebands and the CHASE H$\alpha$ full‑disk scans, to ensure a reliable cross‑comparison between the two instruments. Second, the jet length should be no less than 20 Mm to ensure a sufficient signal-to-noise ratio for reliable detection and analysis. In addition, events that are too faint, severely affected by nearby bright structures, or lack clear morphological boundaries are excluded. The times and locations of all 70 events are listed in Table \ref{table2} in the Appendix. In addition, for the analysis of one particular event, we also utilize observations from the Interface Region Imaging Spectrograph (IRIS; \citealt{Pontieu2014}). IRIS provides high-resolution slit‑jaw images and Si IV spectral line measurements of the jet, with a spatial resolution of about 0.33$^{\prime\prime}$, allowing us to diagnose the plasma dynamics in the transition region.

The primary quantities in our study are the calculated lengths and widths of the jets observed in AIA and CHASE. The methodology for computing these quantities is described as follows. We take axial and transverse slices of the jets, as illustrated in Figure \ref{fig1}(b) for event E1. The temporal resolution of AIA is 12 seconds, while CHASE in full-disk scan mode has a cadence of 60 seconds. To avoid errors in the jet's length and width caused by selecting different observation times, we first determine the time when the jet reaches its maximum height in CHASE and calculate its length and width at that moment. Then, we match this time with the nearest AIA observation time to ensure that the lengths and widths calculated from both wavebands are obtained under temporally aligned conditions.

For each event, we compute the jet length and width on five axial and five transverse slices, average the five values to obtain the final length/width, and adopt the half‑width of the 90\% confidence interval as the uncertainty. Regarding the intensity slicing method, we adopt the intensity slicing method used by \cite{Shimojo1996}, which defines the jet boundary as the point where the intensity drops to 1\% of its peak value. However, because the peak intensity varies substantially across different jets, this fixed-threshold approach can introduce biases: jets with high peak intensities tend to yield artificially short lengths, while those with low peak intensities result in overestimated lengths. To address this, 
we first subtract the pre-event background intensity for each event, to remove the contribution from the ambient corona. We then apply min-max normalization to the intensity profile of each jet before thresholding. We then identify the longest continuous segment where the normalized intensity exceeds 0.05, and define its spatial extent as the length or width (see Figures \ref{fig1}(e) and \ref{fig1}(h)). This threshold is chosen because it effectively removes random noise while preserving the jet structure.

Since jet material is in constant motion, we cannot simply rely on monochromatic images to determine the presence of jet material in the H$\alpha$ waveband. Jet material exhibits pronounced absorption in H$\alpha$ spectra. This feature allows for the tracking of jet material and can be used to obtain its plane-of-sky (POS) and line-of-sight (LOS) velocities as shown in \cite{Huang2025}. Figure \ref{fig1}(c) displays the H$\alpha$ spectral line profile at the green point indicated in Figure \ref{fig1}(b). A significant absorption feature is evident in the blue wing of the line profile. After subtracting the background profile (obtained by averaging the line profiles within the white box in Figure \ref{fig4}(a)), a distinct minimum remains in the blue wing. This minimum reflects the location of maximum absorption by the jet material in the H$\alpha$ line. Therefore, it can reflect the characteristics of the jet material observable in the H$\alpha$ waveband.

 Based on this, we compute the minimum net intensity for each point along the slices. The results for one axial and one transverse slices are shown in Figures \ref{fig1}(f) and \ref{fig1}(i), respectively. We employ the same normalized 5\% intensity method as used for the AIA wavebands to calculate the jet length and width in H$\alpha$. The key distinction is that we now identify the longest continuous region where the value falls below 0.05, defining the spatial distance between the start and end points of this region as the jet length or width. This method enables dynamic tracking of the jet's position and captures the jet material in H$\alpha$ more effectively compared to monochromatic images, which only reflect material at a single wavelength.

For our statistical analysis, we select AIA 304 $\mathrm{\AA}$ for comparison with CHASE H$\alpha$ in terms of jet lengths and widths, because among the various AIA bands, the 304 $\mathrm{\AA}$ images are more suitable for observing filaments and jets and can also avoid obstruction by some hot coronal loops. To verify whether the choice of different AIA bands affects the comparative results, we also calculate the jet lengths and widths in AIA 131 $\mathrm{\AA}$. The comparison between AIA 131 $\mathrm{\AA}$ and AIA 304 $\mathrm{\AA}$ is presented in Figures \ref{fig1}(d) and \ref{fig1}(g). It can be observed that the jet lengths and widths in AIA 131 $\mathrm{\AA}$ and 304 $\mathrm{\AA}$ are very similar. A possible reason for this is that although each AIA waveband has its characteristic temperature, it actually corresponds to a response function, and there is no complete temperature segregation between the different wavebands. This indicates that the choice of AIA waveband to compare with H$\alpha$ does not lead to significant differences in the results. It also demonstrates that studying the temperature structure of jets using only the various AIA bands is far from sufficient, as these bands do not provide fully discrete temperature intervals. In contrast, the H$\alpha$ waveband can only observe cool material and cannot detect the hot material in the corona. This further demonstrates the necessity and importance of our study.

Additionally, we calculate the average velocity and generation time for each jet. For example, Figure \ref{fig4} (c) shows the variation of the jet length observed in AIA 304 $\mathrm{\AA}$ over time. We fit these data with a linear equation $s = k \cdot t + b$, where the obtained coefficient $k$ represents the average velocity of the jet. Furthermore, we determine the onset times of the jets in both AIA and H$\alpha$. When a jet begins to erupt driven by magnetic reconnection, its axial length increases rapidly. By analyzing the temporal evolution of the jet length, we calculate the relative rate of change in jet length between consecutive time frames. A significant increase in jet length (set with a threshold of 10\%) is identified as the onset of the jet eruption. Using the same method, we also compute the onset time of the jet in H$\alpha$ for comparison.

\section{Results}
\label{section3}

\subsection{Statistical Results}
\label{section3.1}

We calculate the generation times of the jets in the AIA 304 $\mathrm{\AA}$ waveband and CHASE H$\alpha$ wavebands. The difference in onset times (H$\alpha$ - 304 $\mathrm{\AA}$) is shown in Figure \ref{fig2}(a). Considering that a full-disk scan by CHASE takes approximately one minute, the estimation error for the onset time in CHASE could be significant. Nevertheless, the maximum temporal offset between two wavebands is only about two minutes. Moreover, the distribution of onset time differences is centered around zero, with no significant delay between the H$\alpha$ and AIA observations. This near simultaneity indicates that the cool components are unlikely to result from the cooling of hot jet plasma; instead, it strongly suggests that both the cool and hot materials are accelerated concurrently.

We also measure the lifetime of each jet following the definition and procedure described in \cite{Liu2019}. The lifetime is defined as the time interval from the jet's first clear appearance to its fallback to the solar surface. Using SDO/AIA 304 $\mathrm{\AA}$ time–distance slices, we identify the start time when the jet structure first becomes observable, and the end time when it can no longer be traced in the time–distance diagrams. Events that move beyond the field of view or lack a clear fallback phase are excluded from the lifetime statistics. A total of 54 jets are reliably measured, with lifetimes ranging from 20 to 89 minutes. The mean lifetime is 45.6 minutes, and the standard deviation is 17.0 minutes.

Figure \ref{fig2}(b) displays the length differences of these 70 jets, calculated using the formula $2(L_{\text{H}\alpha} - L_{304\,\text{\AA}}) / (L_{\text{H}\alpha} + L_{304\,\text{\AA}})$. It can be observed that while the length differences for some jets cluster around zero, others exhibit significant deviations. For most jets, the observed length in H$\alpha$ is smaller than that in 304 $\mathrm{\AA}$. Figure \ref{fig2}(c) shows the width differences of jets, computed as $2(W_{\text{H}\alpha} - W_{304\,\text{\AA}}) / (W_{\text{H}\alpha} + W_{304\,\text{\AA}})$. Similarly, while some differences are near zero, others show substantial deviations, with most jets appearing narrower in H$\alpha$ than in 304 $\mathrm{\AA}$. To simultaneously visualize the differences in both length and width, we present them together in Figure \ref{fig2}(d). The plot reveals that significant differences in length and width do not always occur simultaneously. Based on this, we categorize the jets into four types: Type I – no significant difference in both length and width; Type II – significant difference in width but not in length; Type III – significant difference in length but not in width; Type IV – significant differences in both length and width. To objectively classify these four types, we apply K-means clustering to the data points in Figure \ref{fig2}(d), which automatically divides the data into four groups distinguished by four colors.

Next, we analyze the underlying causes of the four distinct jet morphological types from a statistical perspective. Figure \ref{fig2}(e) shows that the average velocities among the four categories are not significantly different. Therefore, we infer that the observed differences in jet length and width cannot be attributed to variations in jet speed. According to plasma acceleration dynamics, one might expect faster jets to show more pronounced length discrepancies, as cooler, denser material is typically harder to accelerate to the velocities of hotter, lighter plasma. However, our results show no such correlation between jet speed and length/width differences. This implies that the observed morphological variations are not primarily governed by apparent velocity factors, but may be rooted in the jet formation mechanism, as well as intrinsic physical processes such as energy transport and dissipation that govern the cool and hot components.

Furthermore, since Type IV category exhibits significant differences in both length and width, we aim to quantify this difference more precisely. We present the length-to-width ratios ($L/W$) of Type IV jets in both H$\alpha$ and AIA 304 $\mathrm{\AA}$ in Figure \ref{fig2}(f). It can be observed that the $L/W$ ratios in H$\alpha$ are larger than those in 304 $\mathrm{\AA}$ for most cases. Given that both the length and width of jets are smaller in H$\alpha$ than in AIA, we speculate that such jets do not manifest as a uniform plasma column, but rather as a composite structure characterized by a cool core enveloped by a hot shell. The inner low-temperature core traced by H$\alpha$ maintains a compact, short, and narrow morphology under strong magnetic confinement, while the outer high-temperature shell traced by AIA is more likely to expand in the transverse direction due to thermal pressure, appearing more diffuse.

Figure 2(g) shows the proportion of the 70 jets classified into these four categories. Our statistical analysis shows that only 32.9\% of the jets exhibit significant differences in both length and width between EUV and H$\alpha$ observations. In contrast, the majority (67.1\%) of jets show either no significant difference (30.0\%) or a difference in only one dimension (37.1\%). This result challenges the conventional view, which holds that EUV and H$\alpha$ jets should typically possess distinct morphologies and spatial locations. Our analysis, which provides a more complete view of the H$\alpha$ material, leads to a revised conclusion: the material morphologies of the same jet observed in AIA and H$\alpha$ are not systematically different. This contrasts with earlier studies that rely predominantly on H$\alpha$ narrowband images centered on the line center. Since jets are dynamic and their material often shows strong absorption in the line wings, line-core images alone likely present an incomplete picture. Consequently, our complete H$\alpha$ diagnostics reveal that only 32.9\% of jets show significant differences in both length and width.

We further categorize these four jet types based on whether their formation is associated with mini-filaments, with the statistical results displayed in Figures \ref{fig2}(h) and \ref{fig2}(i), respectively. The determination of filament involvement is made by examing CHASE H$\alpha$ and AIA 304 $\mathrm{\AA}$ images prior to jet onset. Events are classified as “with filament” if a distinct mini-filament is clearly visible at the jet's footpoint before eruption and is observed to erupt during the jet event. Events are labeled “without filament” if no clear mini-filament is detected before the jet or if the identified filament does not erupt during the jet process. Among the 70 jet events, 56 show clear mini-filament involvement, which is consistent with previous statistical studies which find that most jets are associated with filaments. Comparing Figures \ref{fig2}(h) and \ref{fig2}(i), it is evident that in events without filament involvement, 50\% belong to Type I (showing no significant difference in either length or width), while the remaining three types collectively account for the other 50\%. This suggests that in the absence of filament, jets are more likely to exhibit consistent morphology across different temperature bands. In contrast, among events with filament involvement, only about 25.0\% are classified as Type I, while the remaining 75.0\% exhibit differences in either length, width, or both, indicating that filament involvement makes differences in length or width more likely to occur. Therefore, it can be inferred that mini-filaments play a role in the jet process, contributing to morphological differences between the cool and hot components. However, questions remain: why do some events without filament involvement still exhibit significant differences in length and width? Conversely, why do some events with filament involvement show consistent length and width across different wavelengths? These questions cannot be fully answered by the statistics alone and will be further explored in Sections \ref{section3.3} and  \ref{section3.4} through detailed analysis of specific case studies. The detailed parameters of the 20 events discussed in the following subsections are listed in Table \ref{tabel1}.

\begin{figure}
   \centering
   \includegraphics[width=0.47\textwidth]{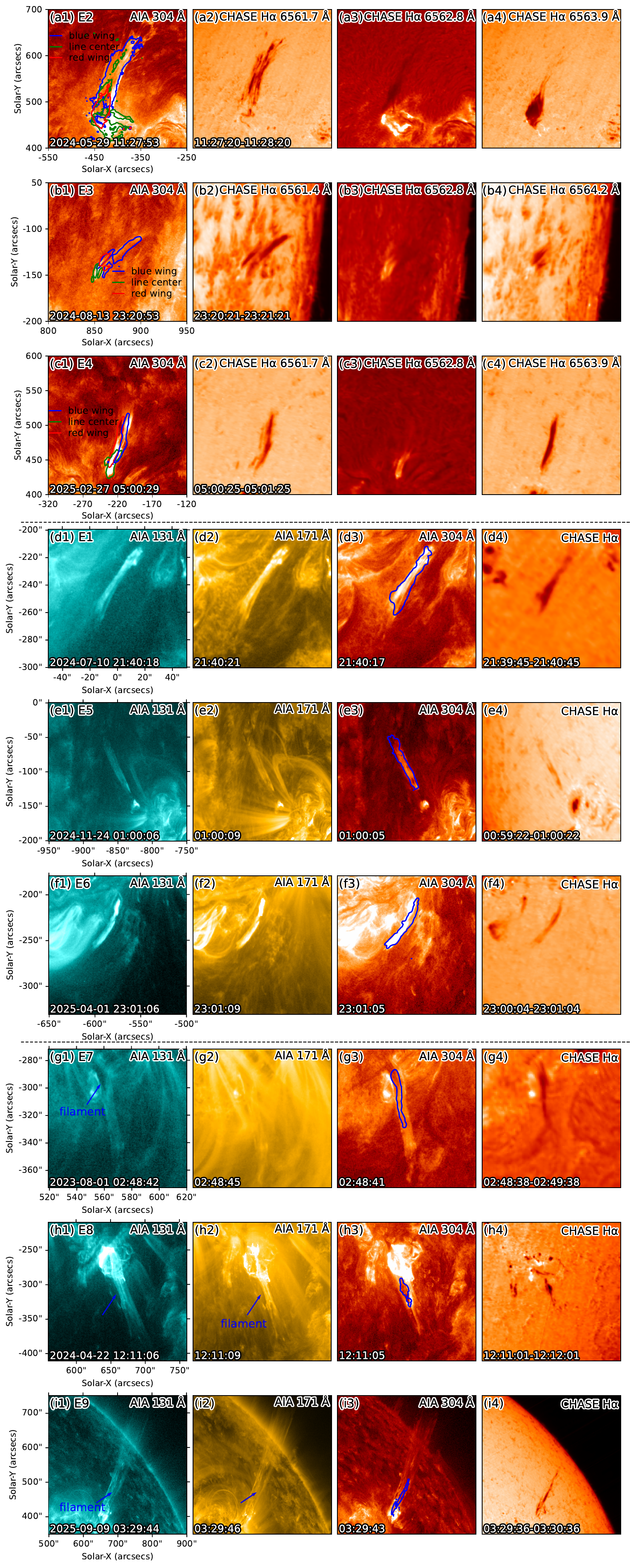}
   \caption{EUV and H$\alpha$ observations of selected jet events. (a)--(c) AIA 304 $\mathrm{\AA}$ image and H$\alpha$ images of three Type I jet events (hereafter E2, E3, and E4, respectively) in the blue wing, line center, and red wing, respectively. (d)--(f) AIA 131 $\mathrm{\AA}$, 171 $\mathrm{\AA}$, 304 $\mathrm{\AA}$, and CHASE H$\alpha$ images for three Type I jet events (E1, E5, and E6). Blue contours represent H$\alpha$ intensity overlaid on the AIA images. (g)--(i) AIA 131 $\mathrm{\AA}$, 171 $\mathrm{\AA}$, 304 $\mathrm{\AA}$, and CHASE H$\alpha$ images for three Type IV jet events with filament involvement (hereafter E7, E8, and E9, respectively). Blue arrows indicate the locations of the filaments. } 
    \label{fig3}
\end{figure}

\subsection{Are We Observing the Same Jet in EUV and H$\alpha$?}
\label{section3.2}
In previous studies, it is widely suggested that the hot and cool components of jets observed in the EUV and H$\alpha$ wavebands, respectively, occupy distinct spatial locations and propagate along different magnetic field lines. However, these conclusions are largely drawn from H$\alpha$ narrowband images near the line center. Given that jets are dynamic structures with continuously evolving velocities and spatial positions, line-core monochromatic images only capture material near the H$\alpha$ line center. Recent studies show that jets often exhibit strong absorption in the blue wing of the H$\alpha$ spectral line, and some even show pronounced absorption in both the blue and red wings (\citealt{Huang2025}; \citealt{Huang2026}). To further illustrate the morphological features of jets in the H$\alpha$ spectral line, we present monochromatic images in the H$\alpha$ blue wing, line core, and red wing for three Type I jets (E2, E3, and E4) in Figures \ref{fig3}(a)--(c). It is clearly evident that jet material is present simultaneously in the blue wing, line core, and red wing, displaying different morphologies. To further confirm the spatial locations of this material, we overlay the contours of the material at these three wavelengths onto AIA 304 $\mathrm{\AA}$ images, as shown in Figures \ref{fig3}(a1), (b1), and (c1), respectively. When combined, the superposition of these three contours forms a composite outline that coincides completely with the jets observed in AIA, sharing identical morphology and spatial characteristics. If only the monochromatic image near the line core (i.e., the green contours) is observed, the jet would appear significantly shorter and narrower than that observed in AIA, and its morphology would clearly differ. Therefore, to capture all cool material observable in the H$\alpha$ waveband, it is essential to combine material from each wavelength. The jet material observed with only one H$\alpha$ narrowband is incomplete. Consequently, conclusions drawn from H$\alpha$ monochromatic or narrowband observations should be treated with caution.

\begin{figure*}
   \centering
   \includegraphics[width=0.95\textwidth]{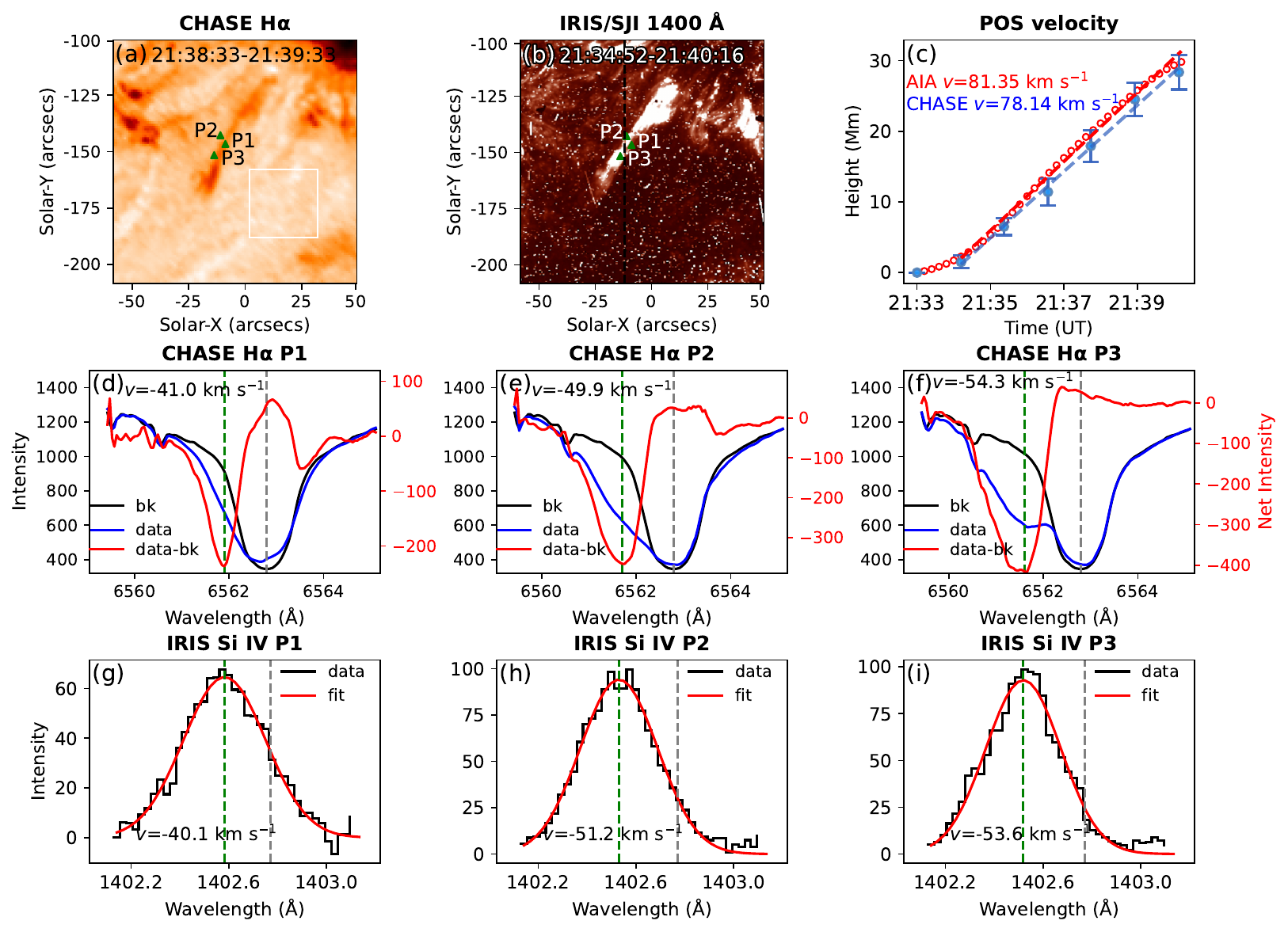}
   \caption{Analysis of the POS and LOS velocities for the cool and hot components in the jet event E1 (2024 July 10). (a) Observational image of this event in CHASE H$\alpha$. The white box marks the region selected for calculating the average background profile. (b) Observational image of this event in IRIS/SJI 1400 $\mathrm{\AA}$. The black dashed line indicates the position of the spectral slit at this time. (c) Temporal evolution of the height measured for this event in CHASE H$\alpha$ and AIA 304 $\mathrm{\AA}$. The dashed lines represent the results of linear fitting. (d)--(f) CHASE H$\alpha$ spectral profiles at positions P1--P3 marked in panel (a). The black, blue, and red lines represent the background profile, the original spectral profile at the point, and the net profile after background subtraction, respectively. The gray dashed vertical line indicates the line-core wavelength $\lambda_0$; the green dashed vertical line indicates the wavelength $\lambda_i$ corresponding to the minimum of the net profile. (g)--(i) IRIS Si IV spectral profiles from P1–P3 marked in panel (b). Black lines indicate the observed profiles, and red lines correspond to the Gaussian fits. The gray dashed vertical line indicates the line-core wavelength $\lambda_0$, the green dashed vertical line indicates the central wavelength $\lambda_i$ obtained from Gaussian fitting.} 
    \label{fig4}
\end{figure*}

To investigate the morphological and dynamical similarities and differences of the same jet in these two wavebands, we conduct some further analysis. Figures \ref{fig3}(d)--(f) present images of three Type I jets (E1, E5, and E6) in three AIA wavebands and CHASE H$\alpha$. The jet morphologies appear nearly identical across the AIA wavebands. Furthermore, by overplotting H$\alpha$ material contours onto the AIA images, we find that the cool material observed in H$\alpha$ aligns closely with the AIA jet morphology in both shape and spatial location. To further investigate the dynamic properties and spectral characteristics of the jets, we select E1 (the same event as in Figure \ref{fig1}(b)) jointly observed by AIA, CHASE, and Interface Region Imaging Spectrograph (IRIS; \citealt{Pontieu2014}), with the analysis results shown in Figure \ref{fig4}. Figure \ref{fig4}(a) displays the CHASE H$\alpha$ image, while Figure \ref{fig4}(b) shows the IRIS slit-jaw imager (SJI) 1400 $\mathrm{\AA}$ image, which captures emission from transition region lines (e.g., C IV and Si IV) formed at temperatures around 10$^5$ K. A comparison reveals consistent jet morphology between these two bands. The H$\alpha$ spectral profiles at three selected locations within the jet are shown in Figures \ref{fig4}(d)--(f), and the corresponding IRIS Si IV line profiles are presented in Figures \ref{fig4}(g)--(i). In the H$\alpha$ profiles, the jet material exhibits a pronounced absorption in the blue wing. After subtracting the background profile (average profile from the white box in Figure \ref{fig4}(a)), a clear minimum remains. We derive the Doppler velocity using the formula $v$$\mathrm{=\frac{c({\lambda}_{i}-{\lambda}_{0})}{{\lambda}_{0}}}$, where $\lambda_i$ is the wavelength shift of this minimum relative to the reference wavelength $\lambda_0$. The value of $\lambda_0$ is determined by applying the 80\% bisector method (\citealt{Zhao2022}) to the average background profile. The Si IV line in the jet also displays significant blueshift (marked by green dashed lines). We apply a single-Gaussian fit to compute its Doppler velocity, using the same formula with $\lambda_0=1402.88$ $\mathrm{\AA}$. At the same three locations, the Doppler velocities derived from H$\alpha$ and Si IV are quite similar, indicating that these two temperature components of the jet share a comparable LOS velocity. Figure \ref{fig4}(c) shows the temporal evolution of the jet length in AIA 304 $\mathrm{\AA}$ and CHASE H$\alpha$. We employ a linear fit $s=k \cdot t+b$, and obtain average POS velocities of 81.35 km $\mathrm{s^{-1}}$ in AIA 304 $\mathrm{\AA}$ and 78.14 km $\mathrm{s^{-1}}$ in CHASE H$\alpha$, indicating that the two temperature components also have similar POS velocities. In summary, both the POS and LOS velocities of the cool and hot components of the jet are remarkably consistent. This demonstrates that the cool and hot components of the jet exhibit nearly identical kinematic characteristics.

\subsection{Physical Origins of Morphological Discrepancies in Filament-Involved Jets}
\label{section3.3}
To investigate the distribution of cool and hot component in jet events involving mini-filaments, we conduct further analysis on this category. As shown in Figures \ref{fig2}(g)--(i), events with filament involvement are more likely to have inconsistent lengths and widths. We present three Type IV events with filament involvement and showing differences in both length and width (E7, E8, and E9) in Figures \ref{fig3}(g)--(i). In the AIA wavebands, clear filament material can be seen inside the jet, and its spatial location coincides with the material observed in H$\alpha$. This suggests that the filament material is enveloped by the hot plasma, leading to the observed differences in length and width. However, except these discrepant events, some events with filament involvement still show no difference between the cool and hot components. Furthermore, there exist events showing differences only in length or only in width. This indicates that discrepancies in length and width might originate from different underlying factors. These questions also need further analysis. Therefore, by examining each individual event with filament involvement, we identify some possible explanations.

\begin{figure*}
   \centering
   \includegraphics[width=0.95\textwidth]{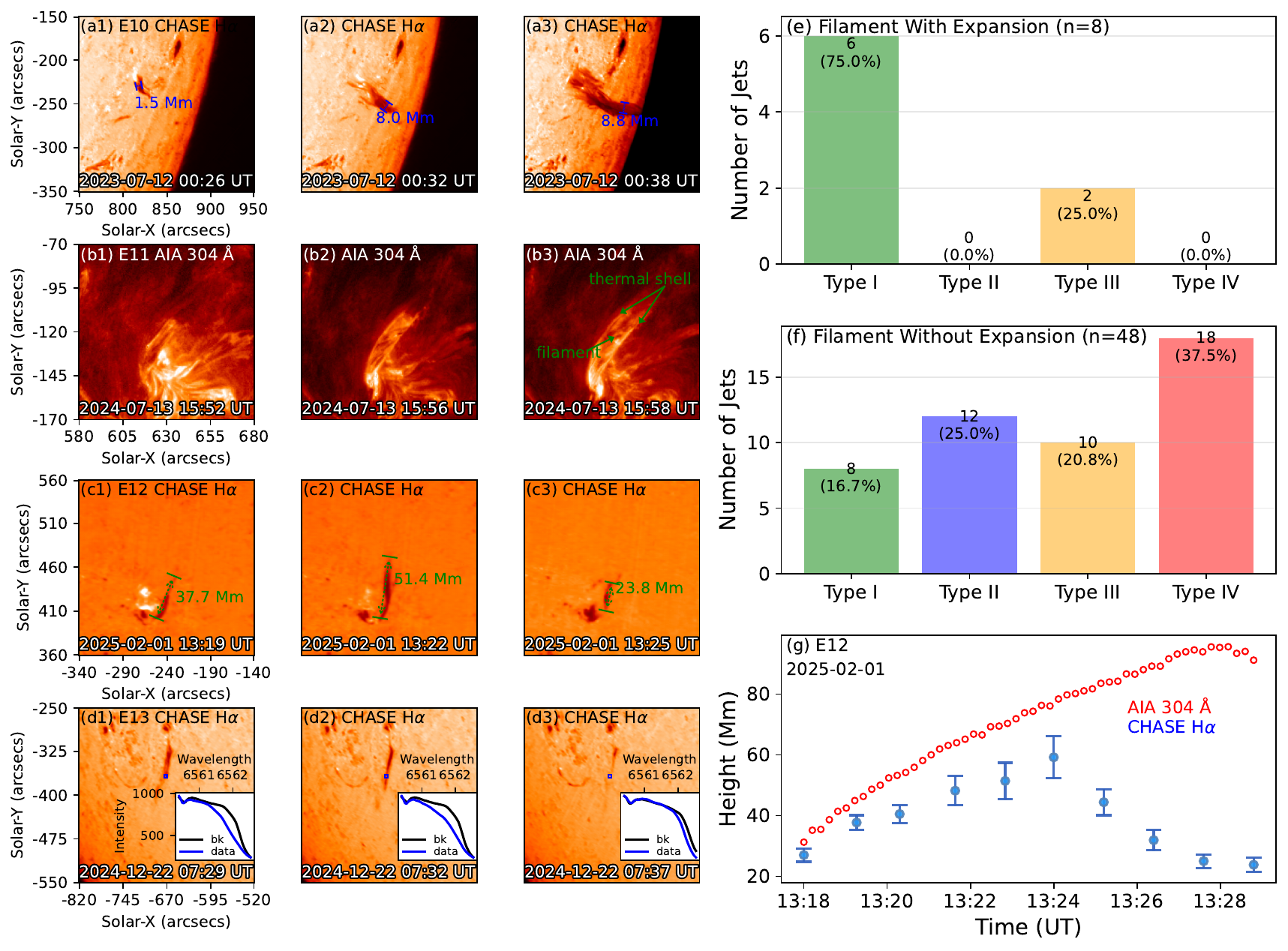} 
   \caption{Analysis of differences in events with filament involvement. (a) Temporal evolution of E10, an event in which the filament undergoes expansion after the eruption. (b) Temporal evolution of E11, an event where the filament is enveloped by a thin hot shell after the eruption. (c) Temporal evolution of E12, an event where the filament's height quits to increase after the eruption. (d) Temporal evolution of E13, an event showing the gradual dispersal of the filament material after the eruption. (e)--(f) Proportions of the four event types with and without filament expansion in filament-involved jets, respectively. (g) Temporal evolution of jet heights measured in AIA 304 $\mathrm{\AA}$ and CHASE H$\alpha$ for E12.}  
    \label{fig6}
\end{figure*}

First, regarding the explanation for width discrepancies, we propose two possible factors: filament expansion and the width of thermal shell. Figure \ref{fig6}(a) shows the temporal evolution of an event with filament involvement but no width discrepancy (E10). It is evident that the filament material undergoes significant expansion during its evolution, with its width increasing from an initial 1.5 Mm to 8.8 Mm. The expansion of a filament during eruption can be attributed to various factors, such as the release of magnetic twist in a flux rope during prominence eruption (\citealt{Xue2021}) or the loss of partial magnetic confinement in previously confined plasma, leading to lateral expansion (\citealt{Wei2024}). In contrast, the three events shown in Figures \ref{fig3}(g)--(i) exhibit no obvious expansion of their filaments during eruption. When a filament expands, the width of the cool material observed in H$\alpha$ matches that of the hot material observed in AIA; whereas in events without filament expansion, the width measured in AIA tends to be wider than that of the cool material in H$\alpha$. We further examine all filament-involved events (n=56) from Figure \ref{fig2}(h) to determine whether significant expansion occurrs during their eruption. The assessment method involves measuring the filament width $W1$ at the initial stage of eruption and $W2$ at the moment when the filament reaches its maximum. If the ratio $W2/W1 > 2$, the filament is considered to have undergone significant expansion. The results are presented in Figures \ref{fig6}(e) and (f). The results indicate that in jet events accompanied by filament expansion, all belong to either Type I or Type III, and no difference in width is observed between their cool and hot components. In contrast, among jet events without filament expansion, the proportions of Type II and Type IV, which do show a width difference, are significantly higher. We therefore conclude that whether a filament expands during the eruption may lead to observable differences in the apparent widths of the cool and hot material. 

The second possible factor is likely the width of the hot shell. If both the cool and hot components are constrained by a strong, confining magnetic structure and do not expand significantly, the resulting hot shell will be thin. Consequently, the widths measured for the cool and hot components in AIA and H$\alpha$ show no significant difference. An example (E11) is provided in Figure \ref{fig6}(b), where a hot shell encloses the central cool filament material. Since the shell is thin, the widths derived from AIA and CHASE H$\alpha$ measurements do not exhibit a notable difference. 

Secondly, regarding the explanation for the differences in length, we propose two possible reasons: insufficient kinetic energy acquired by the cool material, and rarefaction of the cool material. Figure \ref{fig6}(c) shows the temporal evolution of an event with a length discrepancy (E12). The cool material in the jet initially rose, but fell back subsequently. The temporal evolution of the jet length measured by AIA and CHASE, presented in Figure \ref{fig6}(g), indicates that during the initial phase, both the cool and hot materials in the jet are accelerated together. However, in the later stages, the hot material continues to move forward, while the cool material does not. This suggests that although the filament material is accelerated, its higher density may prevent it from acquiring sufficient kinetic energy to maintain the synchronous movement with the hot material. 

Additionally, Figure \ref{fig6}(d) presents another event exhibiting a length discrepancy (E13). Here, the cool material in the jet is initially clearly visible, but over time the top of the jet gradually dissipates. In Figure \ref{fig6}(d), we also display the spectral profiles at the jet apex. It can be seen in Figures \ref{fig6}(d1)--(d2) that the blue wing of the H$\alpha$ line shows significant absorption. However, in Figure \ref{fig6}(d3), the profile in the blue wing is close to the background profile, with only very weak absorption remaining. This further confirms the dispersion of material, leading to the shorter jet length measured in the H$\alpha$ waveband compared to AIA. 

\begin{figure*}
   \centering
   \includegraphics[width=0.95\textwidth]{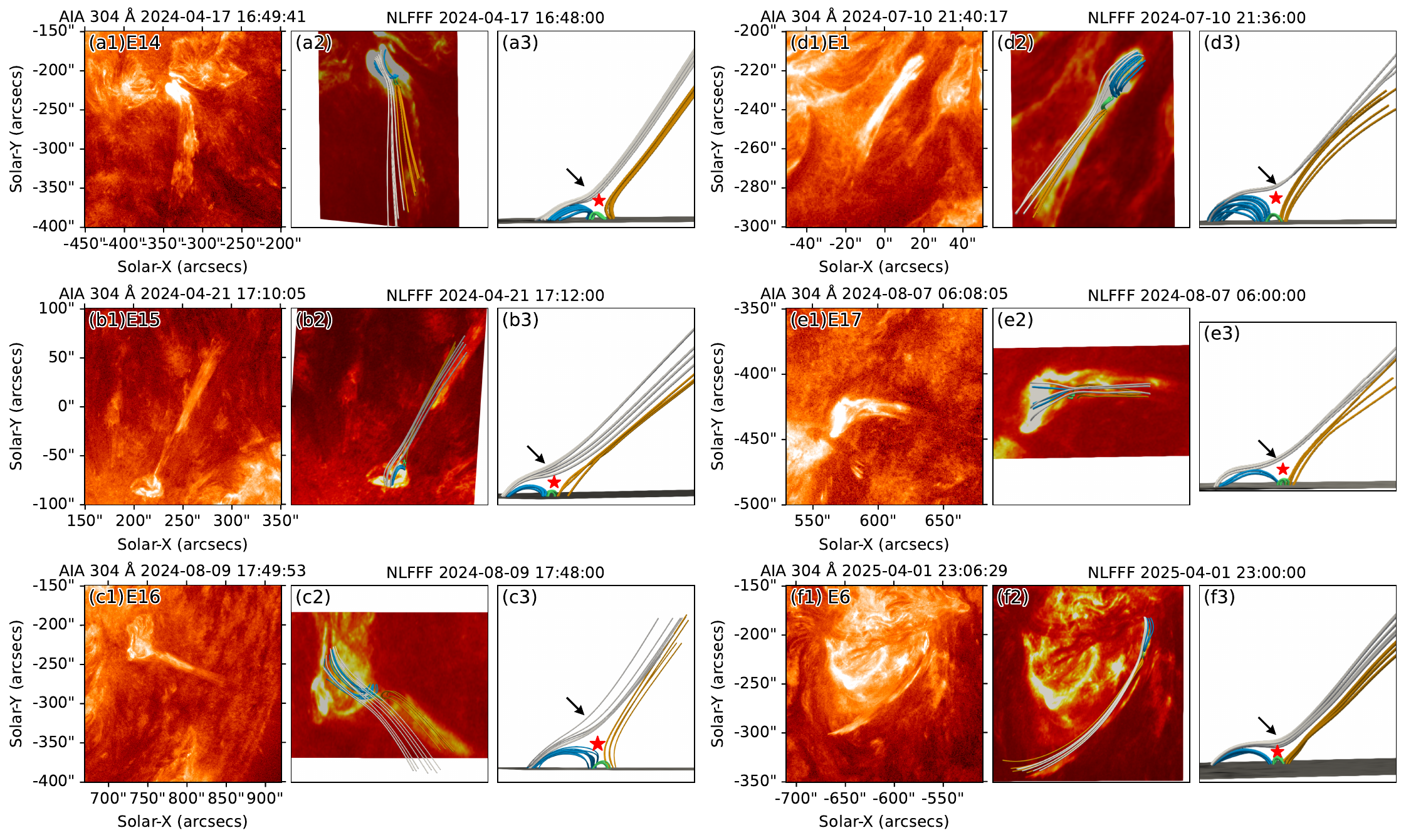}
   \caption{Magnetic field structures obtained from NLFFF extrapolation for six events. (a)--(c) AIA 304 $\mathrm{\AA}$ images, NLFFF extrapolation results viewed from the observational angle, and side views for three Type IV jets without filament involvement (E14, E15, and E16). (d)--(f) AIA 304 $\mathrm{\AA}$ images, NLFFF extrapolation results viewed from the observational angle, and side views for three Type I jets without filament involvement (E1, E17, and E6). The red stars mark the positions of the null point, and the black arrows indicate the locations where the jet spine is most curved.} 
    \label{fig7}
\end{figure*}

\begin{figure}
   \centering
   \includegraphics[width=0.49\textwidth]{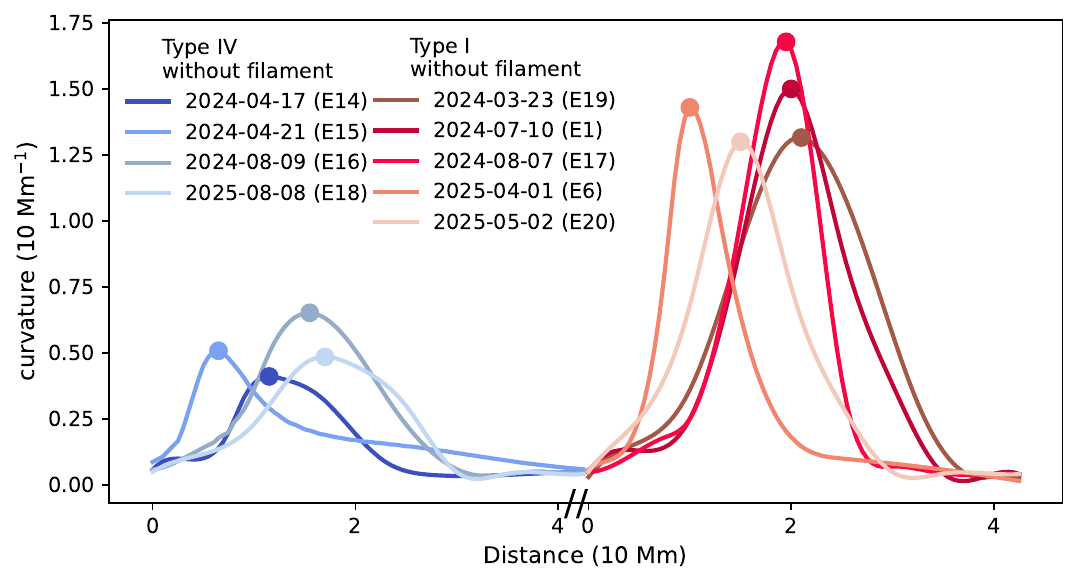}
   \caption{Comparison of magnetic field line curvature versus distance from the jet footpoint for all Type I and Type IV jets without filament involvement (excluding events near the solar limb).} 
    \label{fig8}
\end{figure}

\subsection{Magnetic Field Topology and Its Link to Morphological Differences in Jets without Filament}
\label{section3.4}
Regarding jets without filament involvement, the origin of their cool material is a subject worthy of discussion. Various explanations for the source of this cool material have been proposed. Magnetic tension is still considered the primary energy source for accelerating such material. The cool plasma can be accelerated and ejected by the tension of reconnected magnetic field lines through a “sling-shot” effect (\citealt{Yokoyama1995}).

To compare the similarities and differences in jet events without filament involvement, we select 6 events: three without filament involvement but exhibiting discrepancies in length and width, and three without filament involvement and showing no such discrepancies. To better understand the magnetic field topology, we apply a nonlinear force-free field (NLFFF) extrapolation method to the photospheric vector magnetogram data obtained by Helioseismic and Magnetic Imager (HMI) aboard SDO. We first resolve the 180$^{\circ}$ ambiguity in the transverse components of the vector magnetograms using the minimum energy method. Subsequently, we correct for projection effects following the technique proposed by \cite{Gary1990}, as implemented by \cite{Guo2017}. Next, using the photospheric vector magnetic field of each event as the bottom boundary condition, we extrapolate its potential field in a Cartesian computational domain using the Message Passing Interface Adaptive Mesh Refinement Versatile Advection Code\footnote{\url{https://amrvac.org/}} (MPI-AMRVAC; \citealt{Keppens2012}, \citealt{Porth2014}, \citealt{Xia2018}, \citealt{Keppens2023}). Finally, the potential field is relaxed to a nonlinear force-free field state via the magneto-frictional method (\citealt{Guo2016b, Guo2016a}). Numerous studies have applied the NLFFF extrapolation technique to analyze the magnetic field structure of jets, achieving reliable results, which sufficiently proves the applicability of this method for investigating jets (\citealt{Guo2013}, \citealt{Zhou2025}, \citealt{Huang2025}).

The magnetic field structures of the six events are displayed in Figure \ref{fig7}. The left side shows three Type IV events with discrepancies in both length and width (E14, E15, and E16), while the right side shows three Type I events without such discrepancies (E1, E17, and E6). The second column of each event shows the magnetic field lines projected onto the AIA 304 $\mathrm{\AA}$ images, with the perspective adjusted to a similar viewing angle, facilitating comparison with observations. The comparison shows that the NLFFF extrapolation closely resembles the observed jet morphology and successfully reproduces the magnetic field structure. The third column for each event in Figure \ref{fig7} presents a side view of these open fields. The field lines of Type I events are more curved near the null point, whereas those of Type IV events are relatively straighter. To more quantitatively characterize the degree of field-line curvature, we calculate the curvature of these open fields by $\boldsymbol{K}=\frac{d\boldsymbol{b}}{dl}=\boldsymbol{b} \cdot \nabla \boldsymbol{b}$, with the results shown in Figure \ref{fig8}. We perform the same analysis on all Type I and Type IV events that did not involve filament. However, the number of events without filament involvement is inherently limited, and some events near the solar limb are further excluded. Figure \ref{fig8} shows the curvature as a function of distance for these events. The results show that the peak curvature of Type I events markedly larger than that of Type IV events. Magnetic tension acts as a restoring force of the magnetic field. The relationship between magnetic tension $\boldsymbol{F}$ and curvature $\boldsymbol{K}$ is given by $\boldsymbol{F} = \frac{\boldsymbol{B}^2}{\mu_0} \boldsymbol{K}$. Field lines with smaller curvature correspond to lower magnetic tension. Consequently, weaker magnetic tension may accelerate less cool material, making the cool component more likely to exhibit smaller lengths and widths compared to the hot component. It is important to note, however, that a jet is a highly dynamic process. Conclusions drawn only from a static magnetic field configuration are inevitably incomplete. Nevertheless, this offers a plausible interpretation, based on the current observations, for the differences in length and width observed in jets without filament involvement.

\section{Conclusion and discussion}
\label{section4}
In this section, we summarize our main findings and compare them with previous observational and numerical studies, discussing their implications for future research.

\begin{table*}
\caption{Parameters of the presented 20 events.}
\centering
\resizebox{\textwidth}{!}{%
\begin{tabular}{ccccccccccccccc}
\hline
    & Date       & Longitude & Latitude & velocity\tablenotemark{1} & $t_{304\,\mathrm{\AA}}$\tablenotemark{2} & $t_{H\alpha}$\tablenotemark{3} & $L_{131\,\mathrm{\AA}}\tablenotemark{4}$ & $W_{131\,\mathrm{\AA}}\tablenotemark{5}$ & $L_{304\,\mathrm{\AA}}\tablenotemark{6}$ & $W_{304\,\mathrm{\AA}}\tablenotemark{7}$ & $L_{H\alpha}\tablenotemark{8}$  & $W_{H\alpha}\tablenotemark{9}$ & Filament\tablenotemark{10} & Type\tablenotemark{11} \\
    &            & ($^\circ$)    & ($^\circ$)    & (km s$^{-1}$)   & (UT)     & (UT)       & (Mm)       & (Mm)      & (Mm)       & (Mm)      & (Mm)       & (Mm)     &      &      \\ \hline
E1  & 2024-07-10 & 1        & -13      & 81.4     & 21:34    & 21:34      & 36.9$\pm$3.7   & 5.9$\pm$0.6   & 39.1$\pm$2.2   & 6.8$\pm$0.2   & 35.1$\pm$3.9   & 6.6$\pm$0.2  & N        &I      \\
E2  & 2024-05-29 & -30      & 30       & 197.3    & 11:20    & 11:20      & 157.8$\pm$12.7 & 32.3$\pm$2.0  & 169.4$\pm$4.5 & 26.2$\pm$2.0  & 146.8$\pm$7.5 & 24.3$\pm$1.5 & Y        &I      \\
E3  & 2024-08-13 & 65       & -9       & 78.8     & 23:07    & 23:07      & 57.7$\pm$4.0   & 6.0$\pm$0.3   & 53.1$\pm$4.6  & 6.3$\pm$0.5   & 51.0$\pm$2.2  & 6.0$\pm$1.1  & N        &I      \\
E4  & 2025-02-27 & -16      & 27       & 161.6    & 04:52    & 04:53      & 94.6$\pm$6.6   & 10.8$\pm$0.6  & 90.5$\pm$6.6   & 10.2$\pm$0.5  & 81.9$\pm$4.5   & 11.2$\pm$0.7 & Y        &I      \\
E5  & 2024-11-24 & -60      & -9       & 221.4    & 00:56    & 00:54      & 56.4$\pm$4.4  & 7.3$\pm$0.5   & 52.6$\pm$2.8   & 7.2$\pm$0.5   & 51.4$\pm$5.0   & 6.7$\pm$0.4  & Y        &I      \\
E6  & 2025-04-01 & -37      & -14      & 249.8    & 22:59    & 22:59      & 49.3$\pm$3.7   & 6.3$\pm$0.7   & 48.5$\pm$2.3   & 6.2$\pm$0.5   & 45.6$\pm$3.7   & 6.4$\pm$0.4  & N        &I      \\
E7  & 2023-08-01 & 38       & -18      & 78.3     & 02:41    & No Data          & 49.2$\pm$2.5   & 8.3$\pm$0.6  & 52.0$\pm$4.3   & 7.6$\pm$1.1  & 32.8$\pm$3.8   & 3.5$\pm$0.5  & Y        &IV      \\
E8  & 2024-04-22 & 45       & -16      & 90.7     & 11:58    & 11:58      & 53.1$\pm$4.0  & 9.6$\pm$0.3  & 58.3$\pm$1.8  & 9.4$\pm$0.6  & 18.9$\pm$1.7   & 3.0$\pm$0.2  & Y        &IV      \\
E9  & 2025-09-09 & 50       & 25       & 379.4    & 03:18    & No Data          & 208.9$\pm$13.3 & 24.2$\pm$2.3  & 202.2$\pm$13.0 & 23.5$\pm$1.0  & 102.1$\pm$3.9 & 10.8$\pm$0.5  & Y     &IV      \\
E10 & 2023-07-12 & 64       & -14      & 87.0     & 00:26    & 00:26      & 82.9$\pm$7.0   & 14.8$\pm$1.3  & 76.9$\pm$7.1   & 14.3$\pm$1.4  & 74.7$\pm$9.5   & 12.6$\pm$1.2 & Y        &I      \\
E11 & 2024-07-13 & 40       & -8       & 128.9    & 15:55    & 15:53      & 46.4$\pm$3.9   & 5.0$\pm$0.3  & 50.2$\pm$4.3   & 5.7$\pm$0.7  & 50.5$\pm$2.5   & 5.1$\pm$0.3 & Y        &I      \\
E12 & 2025-02-01 & -17      & 26       & 100.7    & 13:15    & No Data          & 81.8$\pm$4.1  & 4.7$\pm$0.5  & 76.9$\pm$3.8   & 5.4$\pm$0.9  & 51.4$\pm$3.3   & 5.1$\pm$0.7 & Y        &III      \\
E13 & 2024-12-22 & -46      & -18      & 359.9    & 07:15    & No Data          & 94.0$\pm$2.8   & 6.4$\pm$0.4   & 96.8$\pm$7.3   & 5.9$\pm$0.6   & 70.6$\pm$4.7   & 5.0$\pm$0.4 & Y    &III      \\
E14 & 2024-04-17 & -22      & -13      & 154.1    & 16:13    & No Data          & 60.1$\pm$6.0   & 10.7$\pm$0.4   & 57.4$\pm$7.7   & 12.6$\pm$1.8  & 35.8$\pm$3.5   & 5.6$\pm$0.5  & N     &IV      \\
E15 & 2024-04-21 & 13       & -5       & 113.7    & 17:00    & 16:59      & 124.6$\pm$8.4 & 15.7$\pm$1.6  & 122.8$\pm$9.3 & 8.3$\pm$0.8  & 69.6$\pm$5.6  & 5.0$\pm$0.8  & N        &IV      \\
E16 & 2024-08-09 & 54       & -14      & 148.0    & 17:38    & No Data      & 76.3$\pm$6.3   & 9.0$\pm$0.9   & 68.2$\pm$4.3   & 8.3$\pm$1.0  & 33.6$\pm$4.8   & 2.2$\pm$0.4 & N        &IV      \\ 
E17 & 2024-08-07 & 43       & -27      & 134.7    & 05:53    & 05:51      & 22.0$\pm$1.4   & 7.7$\pm$0.6   & 19.7$\pm$2.0   & 7.3$\pm$0.5  & 18.6$\pm$2.1   & 6.9$\pm$0.3 & N        &I      \\
E18 & 2025-08-08 & -32        & 1       & 298.2    & 07:00    & No Data          & 97.3$\pm$8.7 & 16.8$\pm$1.4   & 107.0$\pm$5.0 & 15.5$\pm$0.6  & 96.4$\pm$4.5  & 9.5$\pm$0.3  & Y        &IV      \\
E19 & 2024-03-23 & 4       & -5      & 199.4    & 19:45    & No Data    & 131.4.9$\pm$10.2   & 10.6$\pm$0.6   & 117.7$\pm$11.9   & 10.4$\pm$1.3  & 118.9$\pm$8.6   & 9.6$\pm$1.5 & N        &I      \\
E20 & 2025-05-02 & -32       & 6      & 370.7    & 15:49    & 15:49      & 109.9$\pm$3.5   & 9.7$\pm$0.8   & 103.7$\pm$4.4   & 9.2$\pm$0.7  & 98.3$\pm$5.6   & 8.9$\pm$0.7 & N        &I      \\ \hline
\end{tabular}
}
\tablecomments{1. Average velocity of the jet from AIA 304 $\mathrm{\AA}$; 2. Onset time from AIA 304 \AA; 3. Onset time from CHASE H$\alpha$; 4. Jet length measured in AIA 131 \AA; 5. Jet width measured in AIA 131 \AA; 6. Jet length measured in AIA 304 \AA; 7. Jet width measured in AIA 304 \AA; 8. Jet length measured in CHASE H$\alpha$; 9. Jet width measured in CHASE H$\alpha$; 10. Filament involvement: “Y” indicates presence, “N” indicates no filament involvement; 11. Jet type classification by K-means clustering: Type I - no significant difference in both length and width; Type II - significant difference in width only; Type III - significant difference in length only; Type IV - significant differences in both length and width.}
\label{tabel1}
\end{table*}

\subsection{Summary}
Our study primarily addresses three key questions. The conclusions are summarized below.

Do H$\alpha$ and EUV observations trace the same jet? We conclude that H$\alpha$ and EUV trace different thermal components of a single coherent jet structure. Four lines of evidence support this conclusion: (1) the onset times in AIA and H$\alpha$ are nearly simultaneous; (2) morphological shapes are largely consistent, with only 32.9\% showing discrepancies in both length and width; (3) the two components are co‑spatial, as shown by H$\alpha$ contours overlaid on AIA 304 $\mathrm{\AA}$ images; and (4) their plane‑of‑sky and line‑of‑sight velocities are closely matched, indicating synchronous motion.

Do the cool and hot components show similar or different morphologies? Based on the measured lengths and widths, 30.0\% of the jets show no significant difference in either dimension (Type I); 37.1\% show a difference in only one dimension (Type II or III); and 32.9\% show differences in both dimensions (Type IV). Thus, while the majority (67.1\%) exhibit either full consistency or a difference in only one dimension, a minority (32.9\%) show discrepancies in both.

What distinguishes jets with mini‑filaments from those without? Among the 70 events, 56 (80\%) involve mini‑filament eruptions. Their presence significantly affects the morphological correspondence: only 25.0\% of filament‑involved jets are Type I, compared to 50.0\% of those without filaments. The underlying mechanisms also differ. In filament‑involved jets, the hot outflow envelops the cool filament; discrepancies arise when the filament fails to expand (width) or becomes rarefied or insufficiently accelerated (length), as supported by spatial overlays and case analyses. In jets without filaments, the curvature of the open magnetic field lines plays the dominant role: jets with no morphological discrepancy exhibit a significantly larger peak field curvature than those with discrepancies.

\subsection{Comparison with Observational Studies}
Regarding the relationship between jets and filaments, previous studies suggest that many jets are triggered by the eruption of mini-filaments (\citealt{Sterling2015}; \citealt{Baikie2022}). In our sample, 80\% of jet events involve mini-filaments, further confirming this conclusion. Moreover, multiple studies have provided evidence that the cool component of jets originates from confined, small-scale erupting filaments at the jet base (\citealt{Moore2010}; \citealt{Shen2012}; \citealt{Wang2018}; \citealt{Chen2020}). Our work affirms this conclusion: the cool material observed in H$\alpha$ does indeed correspond to mini-filaments. We further refine this understanding by explaining why jets with filament involvement can exhibit similarities or differences between their cool and hot components, and what factors govern these morphological discrepancies.

For the 20\% of jets without detectable filament eruptions that still exhibit considerable cool material in H$\alpha$, previous work has proposed that the release of cool material may originate from emerging flux carrying chromospheric cold plasma, which is eventually accelerated by the “sling-shot” effect of magnetic tension (\citealt{Yokoyama1996}). Moving beyond the question of its origin, our study focuses on why the morphology of the cool material differs from or resembles that of the hot component. Through NLFFF extrapolation, we find a significant difference in the curvature of magnetic field lines between Type I and Type IV jets (Figure \ref{fig8}), suggesting that magnetic tension, in particular the degree of field-line curvature, plays a crucial role in accelerating the cool material and shaping its morphology.

Previous observations suggest that cool and hot jets are released along different magnetic field lines, with a dynamic connection but asynchronous spatiotemporal evolution (\citealt{Schmieder1994}; \citealt{Alexander1999}). In our study, however, through multiple analyses and more advanced H$\alpha$ material tracking methods, we demonstrate that for some jets, the cool and hot materials not only share the same spatial location but also exhibit similar kinematic characteristics. The complete synchronization of their LOS and POS velocities further indicates consistency in their spatiotemporal evolution. At the same time, we emphasize that the cool material in jets cannot be reliably identified based solely on H$\alpha$ monochromatic or narrow-band images, as a jet is a dynamically evolving and moving structure.

\subsection{Comparison with Numerical Simulations}
Simulation studies of jets also suggest that the release of cool and hot material originates from entirely different spatial locations, with differences observed in their lengths and widths (\citealt{Nishizuka2008}; \citealt{Yokoyama1996}). However, it must be noted that whether jets are triggered by the eruption of magnetic loops or flux ropes, these simulations typically set the background field as a globally open field across the entire domain. This can result in a magnetic structure for the jet that lacks confinement. In the real solar environment, an open field does not refer to a truly open magnetic structure in a global sense, but rather to field lines that open within the field of view. These field lines may indeed be closed globally, but their far footpoint lies outside the simulated or observed region (\citealt{Shen2021}). Precisely because of this, the open field of a jet is not pervasive throughout the domain but is confined, as shown in Figure \ref{fig7}. Therefore, if the simulated open field fills the entire domain, it fails to properly constrain the jet's length and width, potentially leading to uncontrolled dimensions for the cool and hot components. For instance, in the simulation of \cite{Yokoyama1995}, the cool material can appear significantly wider than the hot material—a scenario not observed in our statistical analysis. Consequently, we argue that to further analyze the cool and hot components in jets, MHD simulations based on observed magnetograms are essential. Such simulations would yield magnetic structures closer to those of real jets and significantly advance our understanding of the underlying physics.

\subsection{Implications for Future Automated Detection}
In recent years, machine learning methods have significantly advanced the automatic identification of solar coronal jets. Using deep learning and traditional machine learning models, several studies achieve efficient and accurate detection of coronal jets in EUV wavelengths such as SDO/AIA 304 $\mathrm{\AA}$. For instance, \cite{Liu2024} propose an automated jet identification algorithm based on U-NET, which attains a detection precision of 0.81 at the pixel level, markedly outperforming conventional semi-automatic methods. \cite{Chierichini2025} employ a random forest model that integrates multiple morphological features and successfully identify thousands of new jet candidates from SDO/AIA 304 $\mathrm{\AA}$ data, with a high true-positive rate after manual verification. These works demonstrate that machine learning can provide large-scale, high-quality sample catalogs for jet studies. However, existing automated detection methods mainly focus on EUV wavelengths. To systematically investigate the hot and cool components within jets, it is essential to develop corresponding automated detection approaches for H$\alpha$ observations. \cite{Zheng2024} already implement an automated detection and tracking system for solar filaments in CHASE/H$\alpha$ data, laying a foundation for extending the methodology to jet identification. Future efforts that enable simultaneous automatic identification of jets in both H$\alpha$ and EUV channels could establish a large multi-wavelength jet sample database. Comparative analysis of such data would offer a more comprehensive understanding of the distribution and evolution of hot and cool material in jets. Moreover, current automated identification studies predominantly focus on limb jets, while on-disk jets receive less attention. Numerous jets also exist on the solar disk (see Figure \ref{fig1}(a)), and H$\alpha$ observations often reveal their associated cool plasma structures more clearly than in limb observations. Therefore, developing automated jet detection methods for on-disk regions is also an important and urgent direction for future research.

\begin{acknowledgements}
This study makes use of data from the NASA/SDO (AIA and HMI) and the IRIS missions. IRIS is a NASA small explorer operated by LMSAL, with mission operations at NASA Ames Research Center and key contributions to downlink from ESA and the Norwegian Space Centre. The CHASE mission is supported by the China National Space Administration (CNSA). Numerical computation is supported by the High Performance Computing Center (HPCC) at Nanjing University. Funding is provided by the National Key R\&D Program of China (Grants 2022YFF0503004, 2021YFA1600504, 2020YFC2201201), the National Natural Science Foundation of China (Grants 12333009, 12127901), and the Fundamental Research Funds for the Central Universities (KG202506).
\end{acknowledgements}

\appendix

\begin{table*}[t]
\centering
\caption{Times and locations of all 70 events.}
\setlength{\tabcolsep}{4pt}  
\begin{tabular}{cccc cc cccc}  
\hline
Date & Time (UT) & Longitude ($^\circ$) & Latitude ($^\circ$) & & &
Date & Time (UT) & Longitude ($^\circ$) & Latitude ($^\circ$) \\
\hline
2023-05-25	&14:42	&6	&-25	&&&2024-11-04	&04:58	&12	&14\\
2023-06-25	&22:05	&60	&29	&&&2024-11-06	&08:48	&59	&16\\
2023-07-12	&00:26	&64	&-14	&&&2024-11-24	&00:56	&-60	&-9\\
2023-07-23	&08:28	&51	&4	&&&2024-12-22	&07:15	&-46	&-18\\
2023-08-01	&02:41	&38	&-18	&&&2024-12-24	&13:12	&-16	&-18\\
2023-09-22	&17:14	&31	&21	&&&2025-01-04	&03:47	&-16	&-12\\
2023-11-20	&10:11	&-85	&28	&&&2025-01-18	&02:59	&39	&11\\
2023-12-01	&19:38	&-82	&50	&&&2025-01-31	&10:31	&-33	&24\\
2024-01-10	&10:05	&75	&23	&&&2025-02-01	&13:15	&-17	&26\\
2024-02-05	&04:19	&-71	&-10	&&&2025-02-27	&04:52	&-16	&27\\
2024-03-23	&13:40	&0	&-1	&&&2025-03-01	&02:40	&11	&28\\
2024-03-23	&19:45	&4	&-5	&&&2025-03-25	&14:49	&-89	&-16\\
2024-03-25	&06:43	&25	&-7	&&&2025-03-31	&16:30	&56	&26\\
2024-03-30	&02:09	&-78	&29	&&&2025-04-01	&22:59	&-37	&-14\\
2024-04-13	&05:00	&-83	&-12	&&&2025-04-08	&07:30	&-54	&14\\
2024-04-17	&16:13	&-22	&-13	&&&2025-04-25	&01:57	&27	&17\\
2024-04-21	&17:00	&13	&-5	&&&2025-05-02	&15:49	&-32	&6\\
2024-04-22	&11:58	&45	&-16	&&&2025-05-16	&22:40	&-24	&19\\
2024-05-09	&11:54	&40	&-14	&&&2025-06-11	&02:34	&-89	&23\\
2024-05-14	&16:52	&86	&-16	&&&2025-06-25	&12:40	&30	&-14\\
2024-05-29	&11:20	&-30	&30	&&&2025-07-12	&03:38	&-75	&-16\\
2024-06-08	&03:42	&80	&-20	&&&2025-07-30	&11:27	&48	&-30\\
2024-06-13	&14:55	&-58	&-17	&&&2025-08-08	&04:26	&-76	&2\\
2024-06-17	&10:43	&13	&-27	&&&2025-08-08	&07:00	&-32	&1\\
2024-07-10	&21:34	&1	&-13	&&&2025-09-09	&03:18	&50	&25\\
2024-07-13	&15:55	&40	&-8	&&&2025-09-23	&20:13	&-33	&-12\\
2024-08-07	&05:53	&43	&-27	&&&2025-10-05	&23:40	&60	&9\\
2024-08-09	&17:38	&54	&-14	&&&2025-10-18	&07:12	&64	&7\\
2024-08-13	&23:07	&65	&-9	&&&2025-10-18	&21:06	&89	&30\\
2024-08-27	&08:10	&23	&-33	&&&2025-11-07	&15:48	&-15	&26\\
2024-09-02	&11:16	&35	&-21	&&&2025-11-16	&07:49	&88	&28\\
2024-09-03	&01:40	&-70	&-24	&&&2025-12-04	&15:54	&8	&-18\\
2024-09-10	&05:35	&40	&-24	&&&2025-12-09	&01:35	&40	&-20\\
2024-10-15	&05:28	&29	&-9	&&&2026-01-03	&07:33	&35	&-4\\
2024-10-28	&16:07	&-10	&-21	&&&2026-01-15	&18:43	&-56	&23\\

\hline
\end{tabular}
\label{table2}
\end{table*}

\bibliography{manu}
\bibliographystyle{aasjournalv7}

\end{document}